\documentclass[preprint2]{aastex}

\newcommand{\kms}{km\,s$^{-1}$}

\newcommand{\HII}{\hbox{H~{\sc ii}}}

\begin{document}

\title{THE ENVIRONMENT OF THE STRONGEST GALACTIC METHANOL MASER}

\author{A.\,Sanna\altaffilmark{1}, K.\,M.\,Menten\altaffilmark{1}, C.\,Carrasco-Gonz\'alez\altaffilmark{1, 2}
M.\,J.\,Reid\altaffilmark{3}, S.\,P., Ellingsen\altaffilmark{5}, A.\,Brunthaler\altaffilmark{1},
L.\,Moscadelli\altaffilmark{4}, R.\,Cesaroni\altaffilmark{4}, V.\,Krishnan\altaffilmark{5}.}

\email{asanna@mpifr-bonn.mpg.de}
\altaffiltext{1}{Max-Planck-Institut f\"{u}r Radioastronomie, Auf dem H\"{u}gel 69, 53121 Bonn, Germany}
\altaffiltext{2}{Centro de Radioastronom\'{\i}a y Astrof\'{\i}sica, Universidad Nacional Aut\'onoma de M\'exico, Apartado Postal 3-72, 58090,
Morelia, Michoac\'an, Mexico } 
\altaffiltext{3}{Harvard-Smithsonian Center for Astrophysics, 60 Garden Street, Cambridge, MA 02138, USA}
\altaffiltext{4}{INAF, Osservatorio Astrofisico di Arcetri, Largo E. Fermi 5, 50125 Firenze, Italy}
\altaffiltext{5}{School of Mathematics and Physics, University of Tasmania, Private Bag 37, Hobart, Tasmania 7001, Australia}

\begin{abstract}

The high-mass star-forming site G009.62+00.20\,E  hosts the 6.7\,GHz methanol maser source with the greatest flux density
in the Galaxy which has been flaring periodically over the last ten years. We performed high-resolution astrometric
measurements of the CH$_3$OH, H$_2$O, and OH maser emission and 7~mm continuum in the region. The radio continuum
emission was resolved in two sources separated by 1300~AU. The CH$_3$OH maser cloudlets are distributed along two
north-south ridges of emission to the east and west of the strongest radio continuum component. This component likely
pinpoints a massive young stellar object which heats up its dusty envelope, providing a constant IR pumping for the
Class~\textrm{II} CH$_3$OH maser transitions. We suggest that the periodic maser activity may be accounted for by an
independent, pulsating, IR radiation field provided by a bloated protostar in the vicinity of the brightest masers.
We also report about the discovery of an elliptical distribution of CH$_3$OH maser emission in the region of periodic
variability.

\end{abstract}


\section{INTRODUCTION}

G009.62+00.20 is a star-forming complex located at a parallax distance of 5.2~kpc from the Sun
\citep{Sanna2009,Sanna2014}. Its radio continuum source ``E'', named by \citet{Garay1993}, is the most
compact component among a number of \HII\ regions detected in the field, which align along a ridge of
0.5~pc in size (e.g., \citealt{Liu2011}). Close to G009.62+00.20\,E, the presence of an
embedded, massive, young stellar object (YSO), is inferred through interferometric observations of
high-excitation CH$_3$CN and CH$_3$OH lines, typical hot molecular core tracers, and continuum dust
emission at 2.7~mm \citep{Hofner1996a,Hofner2001}. Source E has not been detected at near-infrared
\citep{Persi2003} or mid-infrared \citep{Linz2005} wavelengths, which argues for an
early stage of stellar evolution. Its continuum emission at 1.3~cm is unresolved with a beam size of
about $0\farcs3$ \citep{Testi2000,Sanna2009}, which sets an upper limit to the physical extent of the
\HII\ emission of $\lesssim0.01$~pc.

In proximity of source E, it is observed several Class~\textrm{II} CH$_3$OH maser transitions (e.g.,
\citealt{Phillips1998,Minier2002}), including the strongest maser emission at 6.7\,GHz  observed in
the Galaxy (5000-6000\,Jy, \citealt{Menten1991}), together with H$_2$O \citep{Hofner1996b}, ground
state OH \citep{Fish2005}, and a rare maser action in the NH$_3$\,(5,5) inversion line \citep{Hofner1994}.
An interesting characteristic of this object is the periodic variability of its Class~\textrm{II} CH$_3$OH masers
\citep{vanderWalt2009}, with a cycle of 244 days over the 10 years monitoring performed so far \citep{Goedhart2014}.

In order to explain this intriguing behavior, it is crucial to image both the maser distribution and the \HII\ region morphology with
high accuracy. For this purpose, we have performed astrometric observations toward G009.62+00.20 of the 6.7~GHz CH$_3$OH, 1.665 and
1.667~GHz OH, and 22.2~GHz H$_2$O masers with the Australian Long Baseline Array (LBA) and the Very Long Baseline Array (VLBA).
Very Long Baseline Interferometry (VLBI) observations were performed in phase-referencing mode for the primary goal of absolute position
calibration at milliarcsecond accuracy. We also conducted Karl\,G.\,Jansky\,Very\,Large\,Array (JVLA) observations in the most extended
configuration at Q band, in order to accurately locate the \HII\ region with respect to the maser species and determine its brightness
distribution.

\section{OBSERVATIONS}

\subsection{MASERS}

We employed the VLBA to observe the $6_{16}-5_{23}$ H$_2$O maser transition (at 22.235079~GHz)
and both main-line $^2\Pi_{3/2}$ $\rm J=3/2$ OH maser transitions (at 1.665401~GHz and 1.667359~GHz).
The LBA was used to observe the $5_{1}-6_{0}A^+$ CH$_3$OH maser transition (at 6.668519~GHz).
At L (1.5\,GHz) and C (6\,GHz) band, we observed in dual circular polarization employing two different frequency
setups for continuum and line measurements. The continuum calibrators were observed using a low number of spectral
channels distributed over a wide bandwidth of 16~MHz and 32~MHz for L and C band, respectively.
Each maser transition was recorded with a 2~MHz wide band with 2048 spectral channels. The receiver setup for
the K (22\,GHz) band observations is described in detail in \citet{Sanna2010a}. Doppler tracking was performed
assuming an LSR velocity of $+2$~\kms\ \citep{Sanna2009}. Observations and source information are summarized
in Table~\ref{obstab}.

The K band observations were carried out with three geodetic-like blocks placed throughout an 8~hr track, in order to
measure and remove atmospheric delays for each antenna \citep{Reid2009}. The H$_2$O maser emission was observed
together with two extragalactic continuum sources: a strong QSO, J1753$-$1843, from the VCS4 catalog which has
an absolute position accuracy of $\pm2$~mas \citep{Petrov2006}, and a weaker QSO, J1803$-$2030, which was previously
employed in U band observations of the $2_{0}-3_{-1}$E CH$_3$OH maser emission at 12.2~GHz \citep{Sanna2009}, whose position
is accurate to only a few tens of mas.  J1753$-$1843 has an angular offset from the target maser of $3.6\degr$ and was
included in both K and L band observations, whereas J1803$-$2030 is offset by $0.7\degr$ and was included in both K and
C band runs. The VLBA and LBA data were processed with the DiFX software correlator \citep{Deller2007} by using an
averaging time of about 0.5~s for the K band dataset and 2~s for both L and C band data.

Data were reduced with the NRAO Astronomical Image Processing System (AIPS). Amplitude calibration should be accurate to better
than $10\%$. For the VLBA antennas, it was obtained by continuous monitoring of the system temperatures. For the LBA, we made use
of a template spectrum extracted from the Parkes data, in order to calibrate the autocorrelation spectra of the remaining antennas (e.g., \citealt{Reid1995}).
The absolute positions of the H$_2$O and OH maser spots were calibrated by phase-referencing with J1753$-$1843.
The H$_2$O and CH$_3$OH maser positions were then registered by using J1803$-$2030 as the absolute position reference.
Maser positions reported in Table~\ref{obstab} are therefore accurate to $\pm2$~mas.

\subsection{RADIO CONTINUUM}

We observed the radio continuum emission in the 7~mm band with the JVLA in its A configuration.
We employed a bandwidth of 256~MHz, in continuum mode and full polarization, for a total run of 1.5~hr.
The bootstrap flux density of the phase calibrator, J1755$-$2232, was 0.29~Jy; the flux scale and
bandpass calibration were obtained on 3C48.
We retrieved VLA B-array data toward G009.62+00.20 at 7~mm from the VLA archive, for the purpose of
recovering extended emission missing in the higher resolution dataset. B-array observations were performed  with a
bandwidth of 100~MHz in two sessions, on 2005 February 18 and April 13, for a total observing time of 2.5~hr.
The phase calibrator, J1820$-$2528, had a bootstrap flux density of 0.57~Jy and 0.64~Jy at the first and second
epochs, respectively; the flux scale and bandpass calibration were obtained on 3C286.

Data reduction was performed with the Common Astronomy Software Applications package (CASA) using standard procedures.
In Table~\ref{obstab}, we report the peak position of the 7~mm continuum emission determined by Gaussian fitting of the
A- and B-array images. In Figure~\ref{maserandradio}, the final images are overlaid with the maser emission in the
region. We further checked the peak position of the radio continuum emission with VLA archival data at X band, observed
between May and June 1998 in A configuration \citep[exp.code AT219]{Testi2000}.
In Figure~\ref{spots}, we registered the peak positions of the three datasets taking into account the proper motion of the
star-forming region \citep{Sanna2009}. These positions agree within about 0\farcs025, suggesting an
astrometric accuracy in the higher resolution dataset at Q band of $\pm 0\farcs025$.

\section{RESULTS}\label{results}

Emissions from different molecular maser species arise within $1''$ from the peak of the radio continuum
component ``E'' (Figure~\ref{maserandradio}). In Table~\ref{masertab}, we list the properties of the
maser emission detected in the channel maps above a 5\,$\sigma$ level. At the spatial resolution of the A array, source
E splits in two sub-components, which hereafter we refer to as E\,1 (southwestern) and E\,2 (northeastern). E\,1 is five
times stronger and offset by about 250~mas (1300~AU) from E\,2. The flux density measured in the B configuration is consistent
with that ($\sim 16$~mJy) reported by \citet{Franco2000} in C configuration, which suggests we recovered all the
significant flux around component E.

OH masers have peak brightness temperatures between $3\times10^{8}$ and $3\times10^{10}$~K and are found further from
both radio continuum peaks than the other masers. Individual spots are slightly resolved with an upper limit in size
of $2\times10^{15}$\,cm. The spot size was obtained by Gaussian deconvolution with the beam.
OH maser cloudlets have FWHM linewidths determined by Gaussian fitting in the range $0.33-0.47$~\kms. Only fits
with FWHM estimated with less than a $10\%$ uncertainty were considered. OH maser spots have line-of-sight velocities
blueshifted by less than 4~\kms\ with respect to the systemic velocity at $+2$~\kms \citep{Hofner1996a}.

Faint water maser emission with peak brightness temperatures in the range $2\times10^{8}-4\times10^{9}$~K is detected
in the region closer to radio component E\,2 than E\,1. Individual spots are slightly resolved with an upper limit in size of
$1\times10^{14}$~cm and arise in four distinct loci along the N-S direction ($\sim 160$~mas). Water maser
cloudlets have linewidths between $0.73-0.95$~\kms\ and emit over a V$_{\rm LSR}$ range of about 13~\kms, including the most
redshifted masing gas from the region at $+11$~\kms.

All the methanol maser emission, except one faint feature, is projected within about 600~AU from the peak of component E\,1
and spans more than three orders of magnitudes in brightness temperature, from $5\times10^{8}$ to $2\times10^{12}$~K.
Individual CH$_3$OH cloudlets are grouped in two N-S filamentary structures to the east and west
of the radio continuum peak E\,1 (Figure~\ref{spots}) and cover a line-of-sight velocity range of 8~\kms. Emission from the strongest
maser spots appears resolved in a core/halo morphology, as it is evidenced in Figure~\ref{mapsum}.
Linewidths range between 0.17 and 0.36~\kms. Emission from the western filament is limited to negative LSR velocities, blueshifted
with respect to the systemic velocity of the region, and its peak brightness temperature is less than
$3\times10^{10}$~K. The eastern filament includes the strongest, Galactic, 6.7~GHz CH$_3$OH maser feature,
which is placed at zero offset in Figure~\ref{spots}.
Its isotropic luminosity of $1\times10^{-4}$~L$_{\odot}$ arises from a cloudlet with an apparent size of $\sim30$\,AU.
Our observing epoch falls in the quiescent period between the 13th and 14th flare cycles on the basis of the ephemeris provided by
\citet{Goedhart2003}. Therefore, both values of brightness temperature and isotropic luminosity have to be taken as lower limits.

We note a quasi-elliptical distribution of CH$_3$OH maser emission which occurs within 150~AU to
the south of the brightest maser feature (Figure~\ref{mapsum}). This emission is clearly detected within a V$_{\rm LSR}$ range
of 0.5~\kms, at redshifted velocities with respect to the brightest maser. The ellipse is best fitted by a major axis of
31\,mas, a minor axis of 13\,mas, and a position angle of $-33\degr$. If the apparent ellipticity
were a projection of a circular ring in Keplerian rotation, viewed from an angle of $\sim25\degr$ from edge-on, the
$0.25$~\kms\ variation over a 16\,mas radius would correspond to a central mass of $\rm 7\times10^{-3}~M_{\odot}$.
The central mass estimated from Keplerian rotation is about eight times that of Jupiter, which rules out a possible
stellar system inside the ring.

\section{DISCUSSION}\label{discuss}

The long-term monitoring performed with the HartRAO telescope by \citet[their Table~2]{Goedhart2014} has set
strong constraints on the periodic activity of the CH$_3$OH masers in G009.62+00.20\,E. At 12.2~GHz, the periodic
flaring has been detected at LSR velocities of 1.25, 1.64, and 2.12~\kms. At 6.7~GHz, the same flaring profile
and recurrence are observed at 1.23, 1.84, 2.24, and 3.03~\kms. In Figure~\ref{spots}, we superpose the 12.2~GHz maser
distribution by \citet{Sanna2009} with that at 6.7~GHz, by aligning the strongest maser features in both transitions.
This alignment provides an optimal match between similar groups of features at 12.2 and 6.7~GHz, which also show LSR velocities
in agreement within about 0.1~\kms. This evidence suggests that both maser lines arise from the same cloudlets.
On the basis of VLBA observations of the 12.2~GHz maser emission during a flare, \citet{Goedhart2005} constrained the
peak position of the flaring activity to within a few tens of mas from the strongest maser feature (at the origin of
Figure~\ref{spots}). No significant flaring delays are detected among the 12.2~GHz features and between features at
similar velocity in both maser transitions \citep{Goedhart2014}. After each flare event, the periodic masers decrease
to the same brightness level they had before the flare, on a time scale of a few months.

As mapped with the LBA (Table~\ref{masertab}), a number of 6.7\,GHz maser cloudlets emit within the velocity range of
the variable peak emission in the single-dish HartRAO spectra. This may suggest that the variable emission arises from
a combination of different, bright, maser cloudlets, as well as possible extended emission resolved out on the long baselines.
These cloudlets should have peak velocities within a maser linewidth from the variable single-dish peaks, in order to contribute
significantly to the emission. By selecting these maser cloudlets in Table~\ref{masertab}, we can confidently confine the region
of periodic CH$_3$OH emission, including 12.2\,GHz masers (\citealt{Goedhart2005}, their animated Figure\,6), inside the solid
box in Figure~\ref{spots}.

In recent years, three possible scenarios have been proposed to explain the periodic variability observed
toward a number of Class~\textrm{II} CH$_3$OH maser sources \citep{Goedhart2014}. The most recent one invokes
stellar pulsation by a OB-type YSO in a pre zero-age-main-sequence (ZAMS) phase, under the assumption of
accretion rates on the order of $\rm 10^{-3}~M_{\odot}~yr^{-1}$ \citep{Inayoshi2013}. This scenario does not apply
to strong, photoionized, radio continuum sources as the model requires effective protostellar temperatures on the order
of 5000~K. Two other models invoke the presence of a binary system of OB-type stars, which eventually provides the periodic
modulation of either 1) the radiation field that is amplified by the maser
\citep{vanderWalt2011}, or 2) the IR radiation field that cause the population inversion of the maser levels \citep{Araya2010}.
Model 1) requires the masing region amplifying a \hbox{H~{\sc ii}} continuum emission in the background, which is
periodically enhanced by a colliding-wind binary \citep[their Figure~1]{vanderWalt2011}.
Model 2) predicts the presence of a circumbinary disk which periodically accretes material onto a binary system, following
simulations by \citet{Artymowicz1996}, and periodically heats up a dusty envelope increasing the IR
radiation field (see also a similar model by \citealt{Parfenov2014}).

\begin{figure*}
\centering
\includegraphics[angle= 0, scale= 0.8]{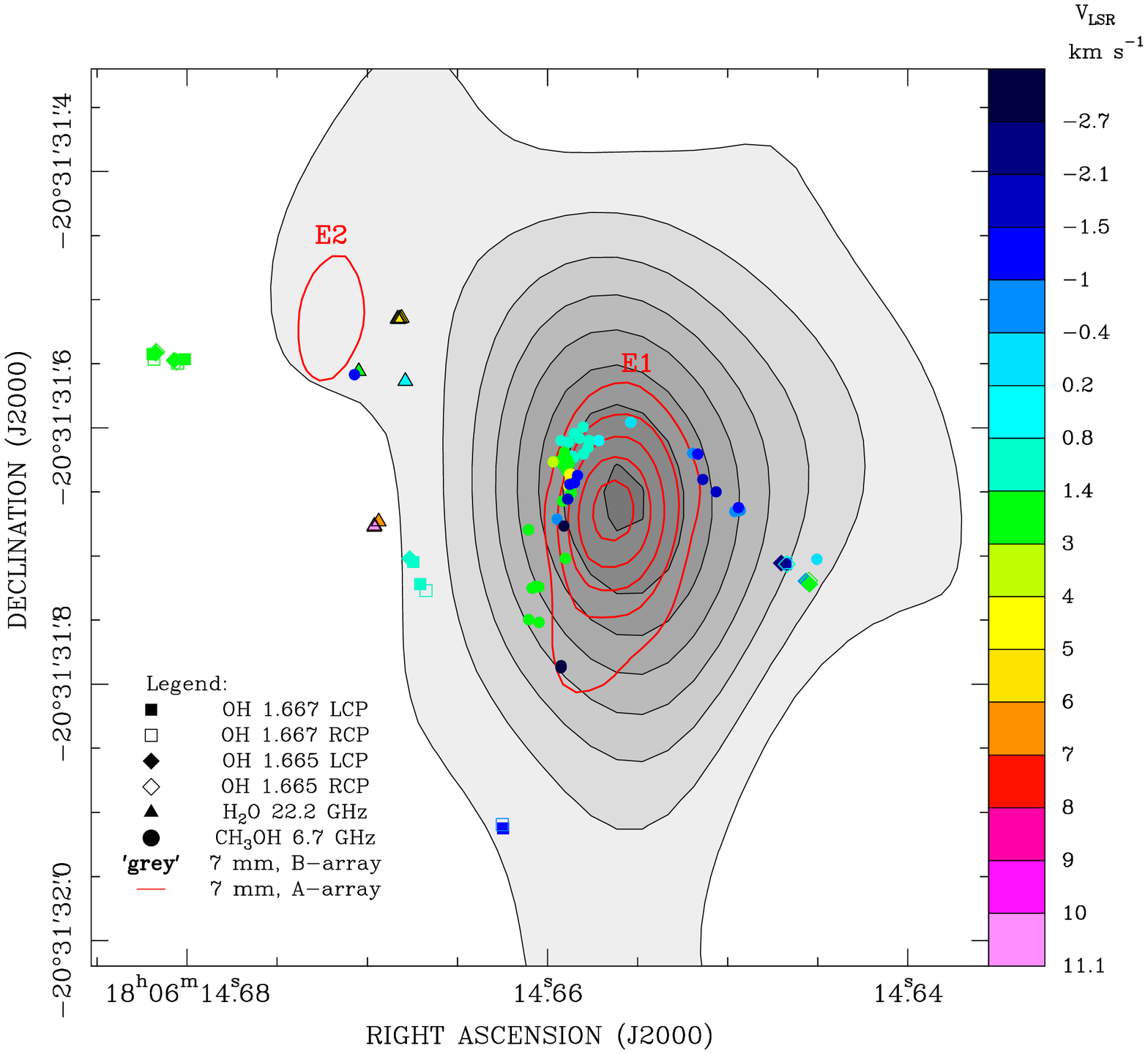}
\caption{Combination of the VLBI measurements of different maser species obtained toward G009.62+00.20\,E.
The distribution of the maser cloudlets is superposed on the VLA maps at 7~mm, obtained
with both the B and A configurations. Linear contouring was used for both maps starting from a 4\,$\sigma$ level (Table~\ref{obstab});
the first two negative levels fall below this threshold. Maser velocities along the line-of-sight are color coded according
to the righthand V$_{\rm LSR}$ scale.
\label{maserandradio}}
\end{figure*}

\begin{figure*}
\centering
\includegraphics[angle= 0, scale= 0.8]{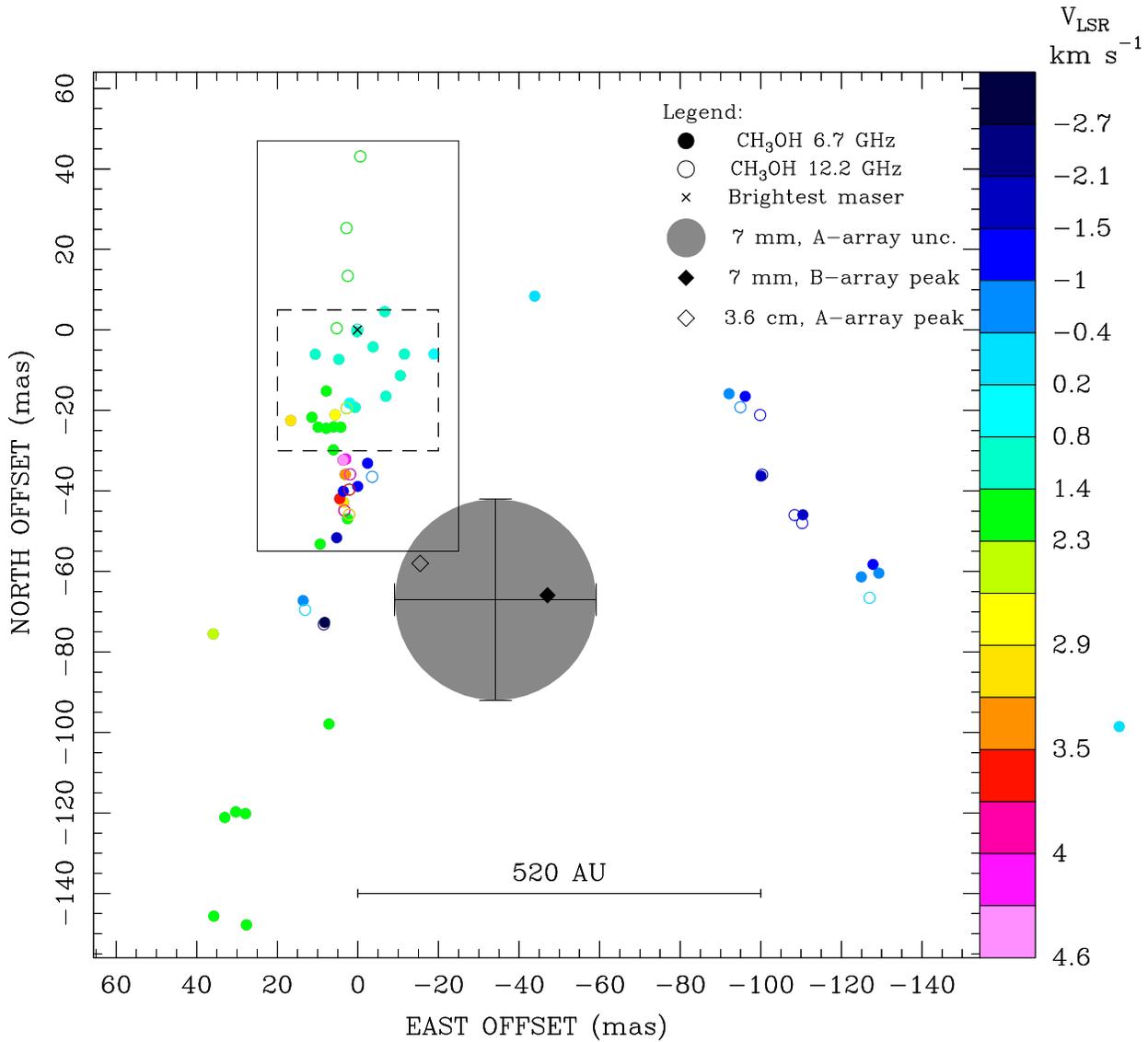}
\caption{Close-up view of the 6.7 and 12.2~GHz \citep{Sanna2009} maser emission arising within $\sim600$~AU from the radio
continuum peak. The 6.7~and 12.2~GHz masers are aligned by superposing their strongest features (cross) at zero offset.
Maser velocities along the line-of-sight are color coded according to the righthand V$_{\rm LSR}$ scale. The large grey
circle gives the astrometric uncertainty of the radio continuum peak. The solid and dashed boxes mark the region of maser
variability, and that studied in Figure~\ref{mapsum}, respectively. Linear scale at the bottom.
\label{spots}}
\end{figure*}

\begin{figure*}
\centering
\includegraphics[angle= 0, scale= 0.6]{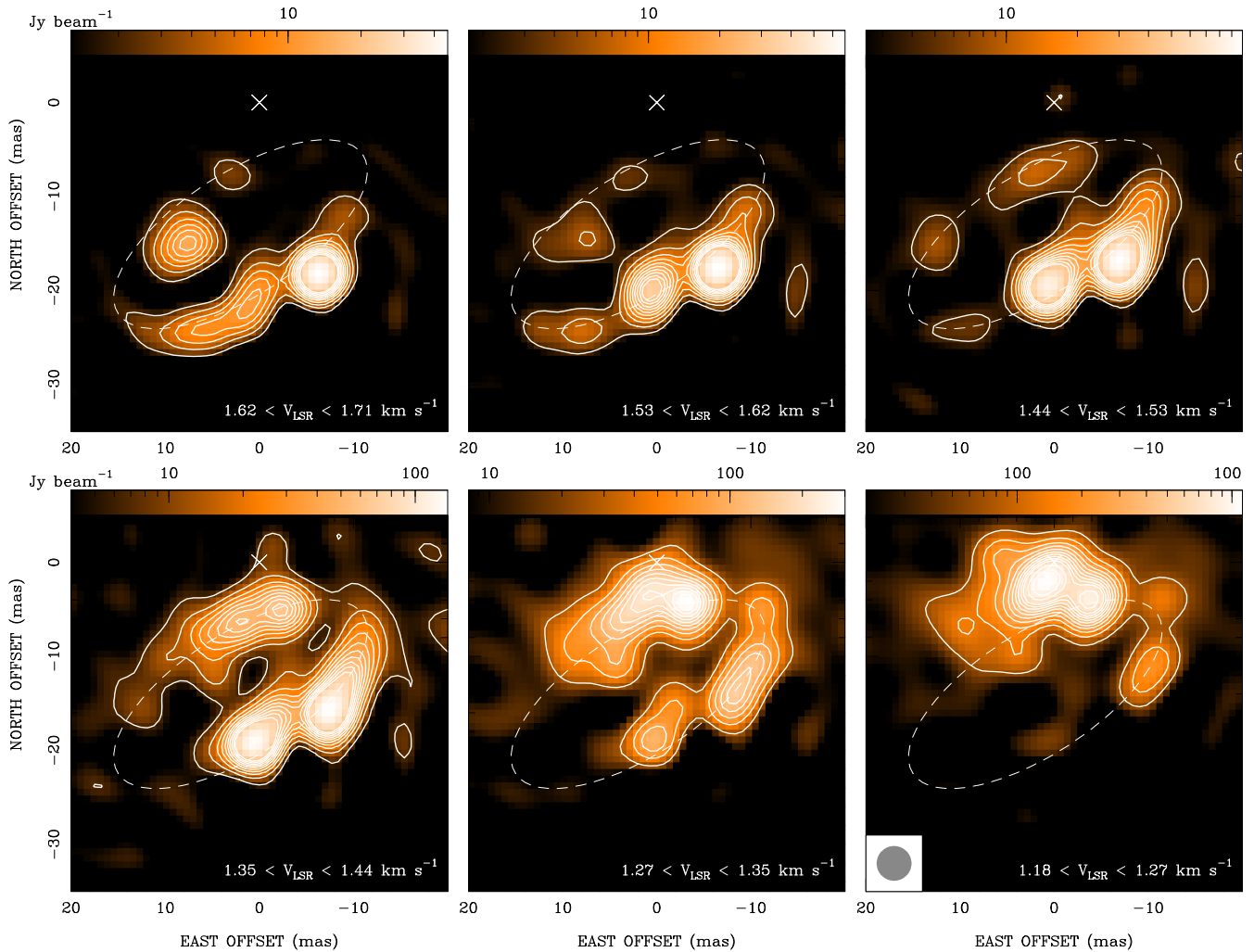}
\caption{6.7~GHz CH$_3$OH maser maps within $\sim150$~AU south of the brightest CH$_3$OH maser (white cross).
Each panel was obtained by summing the maser emission in three, consecutive, velocity channels (ranges in each
bottom right corner). Contour levels start at 5\,$\sigma$ and increase by 5\,$\sigma$ steps for all panels but the two
lower velocity maps which start at 10\,$\sigma$ by 10\,$\sigma$ steps. For each plot, the logarithmic color scale
(upper wedge) range from a 3\,$\sigma$ level up to the peak emission. The dashed ellipse marks the best fit to the overall
maser emission. Restoring beam in the lower right panel.
\label{mapsum}}
\end{figure*}

Currently, we cannot rule out a periodic variation of the radio continuum emission amplified by the maser cloudlets.
If the source of input signal to the maser were coincident with component E\,1, which is projected against the region of maser
variability, maser cloudlets with different LSR velocities than the variable masers would likely lie further away from E\,1.
Otherwise, we make use of the YSO multiplicity, evidenced by the 7\,mm continuum peaks, and the presence of high accretions rates
($\rm \sim 4\times10^{-3}\,M_{\odot}\,yr^{-1}$) recently detected toward G009.62+00.20\,E by \citet{Liu2011}, to suggest
that the maser variability may be explained by the superposition of two, independent, IR sources, which provide
the pumping mechanism for Class~\textrm{II} CH$_3$OH masers \citep{Sobolev1994,Cragg2005}. First, a constant IR
radiation field provided by the warm dusty envelope associated with the radio continuum source E\,1 (Figure~\ref{maserandradio}).
This environment would sustain the population inversion of the CH$_3$OH maser levels over the whole region and cause the maser
emission during the quiescent flaring period.
Second, an IR radiation field periodically enhanced by a growing massive YSO in the vicinity of the eastern CH$_3$OH maser
distribution, following the model by \citet{Inayoshi2013}. This object cannot be the main cause of radio continuum emission
in the region. Instead, we consider an object associated with the faint radio component E\,2, which lies (in projection) close
to ($\sim1000$\,AU) the peak of the CH$_3$OH periodic activity.

To test this hypothesis, we compare the number of photons emitted by a warm dust layer surrounding source  E\,1, with
those provided by the photosphere of a bloated protostar at the position of E\,2. Assuming saturated masers, since the CH$_3$OH
maser features increase by a factor up to 3 during the flaring period \citep{Goedhart2014}, these two sources of
IR photons should be comparable. We consider a dust temperature on the order of 150~K \citep{Cragg2005}, consistent with the
rotational temperature measured in the region from CH$_3$CN and H$_2$CS molecules \citep{Hofner1996a,Liu2011}. For a frequency
of about 13.5\,THz, corresponding to the excitation of the upward links which pump the 12.2\,GHz masers \citep{Sobolev1994}, we
obtain a number of emitted photons per unit time and area (N$_{dust}$) of $\rm 1.7 \times 10^8~s^{-1}~m^{-2}~Hz^{-1}$. We can
parameterize the  number of photons which effectively reaches a masing cloudlet, as a function of the relative distance of the
the dust layer (R$_{dust}$) and the cloudlet (R$_{maser}$) with respect to the exciting YSO, $\rm 1.7 \times 10^8 \times (R_{dust}/R_{maser})^2~s^{-1}~m^{-2}~Hz^{-1}$.
We then consider a secondary, variable, source of IR photons provided by a bloated protostar which pulsates at the period of
the maser flares. Following eq.\,(3) of \citet{Inayoshi2013}, this object would attain a protostellar radius of about
600\,R$_{\odot}$, and a protostellar luminosity consistent with the IRAS fluxes measured in the region (e.g.,
Table\,1 of \citealt{Inayoshi2013}). Such a protostar, with an effective temperature of 5000\,K, would provide a
number of 13.5\,THz photons (N$_{proto}$) of $\rm 7 \times 10^5~s^{-1}~m^{-2}~Hz^{-1}$ at the position of the periodic masers
(box in Figure~\ref{spots}). Providing that the maser is saturated, we can reason on the position of the bloated
protostar with respect to the masing gas. Assuming the bloated protostar at the position of E\,2, the two pumping sources would
be comparable if the masing cloudlets are observed in the foreground with respect to the warm dust layer (e.g.,
R$_{maser} \sim 10 \cdot R_{dust}$). Otherwise, the protostar should lie closer to ($<100$\,AU) the region of periodic
variability for R$_{maser} \sim R_{dust}$.

A critical test for the presence of a pulsating protostar in the vicinity of the radio component E\,2, may be provided by
the eastern cluster of OH masers at a V$_{\rm LSR}$ of 1.65~\kms\ (Figure~\ref{maserandradio}). Since OH masers are
radiatively pumped by an IR radiation field similar to the one exciting the Class~\textrm{II} CH$_3$OH masers
\citep{Cragg2002}, OH maser features close to the source of IR radiation may be subject to a similar periodic flaring
as well, which would rule out model 1).

\acknowledgments

Comments from an anonymous referee are gratefully acknowledged.
This work was partially funded by the ERC Advanced Investigator Grant GLOSTAR (247078).
This work made use of the Swinburne University of Technology software
correlator, developed as part of the Australian Major National Research
Facilities Programme and operated under licence. The Australia Telescope Compact Array
(/ Parkes radio telescope / Mopra radio telescope / Long Baseline Array) is part of the
Australia Telescope National Facility which is funded by the Commonwealth of Australia for
operation as a National Facility managed by CSIRO. The VLBA and JVLA are operated by the National Radio
Astronomy Observatory (NRAO). The NRAO is a facility of the National Science Foundation operated
under cooperative agreement by Associated Universities, Inc.

{\it Facilities:} \facility{VLBA, LBA, JVLA}.






\clearpage

\clearpage

\begin{deluxetable}{cccccccrccrc}
\tabletypesize{\scriptsize}
\rotate
\center
\tablecaption{Observation and Source Information\label{obstab}}
\tablewidth{0pt}
\tablehead{
\colhead{ }         &    \colhead{ } &     \colhead{ }    &    \colhead{ } & \colhead{ }          & \colhead{ }      & \colhead{ }
&        \multicolumn{2}{c}{Peak position}         & \colhead{ }              & \colhead{ } & \colhead{ } \\
\colhead{Telescope} & \colhead{Band} & \colhead{exp.code} &    \colhead{Obs.Date} &  \colhead{Spec.Res.} &
\colhead{HPBW\tablenotemark{b}}   & Image rms &  \colhead{R.A.~(J2000)} & \colhead{Dec.~(J2000)} & \colhead{F$_{\rm peak}$} &
\colhead{F$_{\rm int}$} & \colhead{V$_{\rm LSR}$} \\
\colhead{ }         &    \colhead{ } &     \colhead{ }    &    \colhead{ } & \colhead{(km s$^{-1}$)}  & \colhead{(mas)} &
\colhead{(Jy beam$^{-1}$)} & \colhead{(h m s)} & \colhead{($\degr$ ' '')} & \colhead{(Jy beam$^{-1}$)} & \colhead{(Jy)}
& \colhead{(km s$^{-1}$)} \\}

\startdata
VLBA     & L & BS224A & 2013 Jul 02 & 0.18 & $ 18  \times 18  $  &  0.03 &  18 06 14.6820 & $-$20 31 31.542 &  21.15 &   30.12  &  1.65 \\
LBA      & C & V255A  & 2008 Mar 30 & 0.04 & $ 3.7 \times 3.7 $  &  0.03 &  18 06 14.6586 & $-$20 31 31.604 & 1010.63 & 3894.83   &1.05 \\
VLBA     & K & BS208A & 2012 Jan 19 & 0.21 & $ 0.8 \times 0.8 $  &  0.01 &  18 06 14.6684 & $-$20 31 31.515 &   0.78 &    1.10  & -0.52 \\
        \multicolumn{10}{c}{\textbf{Radio Continuum}} \\
JVLA--A     & Q & AS111   & 2011 Jun 16 & ...          & $ 93 \times 50$    at $-1.5\degr$ & $2\cdot10^{-4}$ &  18 06 14.6562 &
$-$20 31 31.671 & $5.6\cdot10^{-3}$ & $13.6\cdot10^{-3}$ & ...  \\
JVLA--B\tablenotemark{a}  & Q & AG0685 & 2005 Feb/Apr & ...  & $ 284 \times 163 $ at $0.6\degr$  & $2\cdot10^{-4}$ &
18 06 14.6555 & $-$20 31 31.655 & $6.9\cdot10^{-3}$ &   $18.6\cdot10^{-3}$  & ...  \\

\enddata
\tablecomments{Columns~1 and~2 specify the interferometers and radio bands.
Columns~3, 4, and~5 list the experiment code, observational epoch, and spectral resolution employed, respectively.
Columns~6 and~7 give the restoring beam size and the rms noise of the final imaging.
Columns~8 and~9 give the absolute position of the reference maser spots, and the peak position of the continuum emission.
Columns~10 and~11 give the peak brightness and the flux density of the reference spots and the main continuum component
determined by Gaussian fitting. Column~12 reports the LSR velocity of the reference spots.}
\tablenotetext{a}{The flux density was integrated within a box of $1''$ from the peak.}
\tablenotetext{b}{Maser cubes used a circular beam size equal to the geometrical average of the dirty beam.}

\end{deluxetable}

\begin{deluxetable}{lrrrr}
\tabletypesize{\scriptsize}
\tablecaption{Parameters of maser features\label{masertab}}
\tablewidth{0pt}
\tablehead{
\colhead{Feature} & \colhead{V$_{\rm LSR}$} & \colhead{F$_{\rm peak}$}   &  \colhead{$\Delta \rm x$} & \colhead{$\Delta \rm y$} \\
\colhead{\#}      & \colhead{(\kms)}        & \colhead{(Jy beam$^{-1}$)} &  \colhead{(mas)}          & \colhead{(mas)}          \\}

\startdata

   \multicolumn{5}{c}{\textbf{OH features at 1.667~GHz -- LCP} }         \\

1   & $ 1.65$ & 21.15 &  $   0     \pm 0.7 $ & $    0   \pm 0.2  $  \\
2   & $ 2.01$ &  0.68 &  $   -25.7 \pm 0.5 $ & $  -4.2  \pm 1.7  $  \\
3   & $-1.05$ &  0.62 &  $  -273.8 \pm 0.8 $ & $ -370.3 \pm 1.0  $  \\
4   & $ 1.47$ &  0.31 &  $  -203.8 \pm 1.6 $ & $ -162.4 \pm 2.3  $  \\
5   & $ 1.11$ &  0.27 &  $  -209.2 \pm 1.2 $ & $ -179.7 \pm 1.8  $  \\
6   & $-1.59$ &  0.20 &  $  -493.9 \pm 2.7 $ & $ -164.1 \pm 2.1  $  \\

   \multicolumn{5}{c}{\textbf{OH features at 1.667~GHz -- RCP} } \\

7   & $ 1.47$ &  2.06 &  $    -1.5 \pm 1.1 $ & $   -4.2 \pm 0.4  $  \\
8   & $ 0.57$ &  1.39 &  $  -494.7 \pm 1.3 $ & $ -163.1 \pm 0.3  $  \\
9   & $ 2.01$ &  0.53 &  $   -20.1 \pm 0.9 $ & $   -7.4 \pm 0.9  $  \\
10  & $ 1.11$ &  0.39 &  $  -213.7 \pm 1.7 $ & $ -184.8 \pm 2.1  $  \\
11  & $-0.69$ &  0.26 &  $  -273.2 \pm 1.2 $ & $ -367.3 \pm 1.3  $  \\

   \multicolumn{5}{c}{\textbf{OH features at 1.665~GHz -- LCP} } \\

12  & $ 1.65$ & 10.01 &  $    -2.9 \pm 0.6 $ & $    0.6 \pm 0.5  $  \\
13  & $ 1.83$ &  1.06 &  $   -17.4 \pm 2.0 $ & $   -5.1 \pm 2.0  $  \\
14  & $-0.69$ &  0.86 &  $  -510.0 \pm 0.9 $ & $ -177.2 \pm 0.7  $  \\
15  & $-2.32$ &  0.73 &  $  -490.8 \pm 0.5 $ & $ -163.4 \pm 0.8  $  \\
16  & $ 2.01$ &  0.60 &  $  -512.8 \pm 1.0 $ & $ -179.6 \pm 1.8  $  \\
17  & $ 1.29$ &  0.38 &  $  -201.1 \pm 1.3 $ & $ -159.8 \pm 1.3  $  \\

   \multicolumn{5}{c}{\textbf{OH features at 1.665~GHz -- RCP} } \\

18  & $ 1.29$ &  5.00 &  $  -495.1 \pm 2.3 $ & $ -164.1 \pm 1.6  $  \\
19  & $ 2.01$ &  4.35 &  $  -512.3 \pm 1.3 $ & $ -177.1 \pm 1.3  $  \\
20  & $ 2.19$ &  2.73 &  $   -19.1 \pm 0.6 $ & $   -6.2 \pm 0.2  $  \\
21  & $ 1.65$ &  2.17 &  $    -3.4 \pm 0.6 $ & $    1.5 \pm 0.5  $  \\

   \multicolumn{5}{c}{\textbf{22.2~GHz H$_2$O features} } \\

1  & $ -1.99 $ &  1.05 &  $    0.34 \pm 0.07 $ & $   0.03 \pm 0.07 $  \\
2  & $ -0.52 $ &  0.78 &  $     0   \pm 0.05 $ & $    0   \pm 0.03 $  \\
3  & $  3.06 $ &  0.36 &  $   -0.22 \pm 0.15 $ & $  -0.19 \pm 0.12 $  \\
4  & $  6.22 $ &  0.32 &  $   14.99 \pm 0.04 $ & $-158.11 \pm 0.05 $  \\
5  & $  1.80 $ &  0.30 &  $   30.48 \pm 0.03 $ & $ -40.93 \pm 0.07 $  \\
6  & $  7.70 $ &  0.22 &  $   18.43 \pm 0.04 $ & $-162.34 \pm 0.05 $  \\
7  & $ 11.07 $ &  0.17 &  $   18.31 \pm 0.04 $ & $-161.54 \pm 0.09 $  \\
8  & $  4.75 $ &  0.17 &  $   -0.10 \pm 0.08 $ & $  -0.03 \pm 0.12 $  \\
9  & $  6.01 $ &  0.15 &  $   -2.81 \pm 0.11 $ & $   0.67 \pm 0.11 $  \\
10 & $  5.59 $ &  0.11 &  $   -1.50 \pm 0.08 $ & $   0.31 \pm 0.13 $  \\
11 & $  0.54 $ &  0.06 &  $   -5.86 \pm 0.09 $ & $ -48.59 \pm 0.15 $  \\

   \multicolumn{5}{c}{\textbf{6.7~GHz CH$_3$OH features} } \\

  1    &  $  1.05 $  & $ 1010.63 $  &  $   0    \pm  0.54  $  &  $   0    \pm  1.24  $  \\
  2    &  $ -0.45 $  & $  565.03 $  &  $  13.50 \pm  0.30  $  &  $ -67.26 \pm  0.72  $  \\
  3    &  $  0.92 $  & $  474.85 $  &  $  -6.76 \pm  0.25  $  &  $   4.53 \pm  0.54  $  \\
  4    &  $  1.18 $  & $  347.72 $  &  $  -3.86 \pm  0.80  $  &  $  -4.25 \pm  0.36  $  \\
  5    &  $  1.13 $  & $   80.94 $  &  $ -10.63 \pm  0.44  $  &  $ -11.37 \pm  0.32  $  \\
  6    &  $  1.13 $  & $   69.66 $  &  $  10.47 \pm  0.44  $  &  $  -6.07 \pm  0.53  $  \\
  7    &  $  1.31 $  & $   49.27 $  &  $  -7.06 \pm  0.38  $  &  $ -16.51 \pm  0.92  $  \\
  8    &  $  1.31 $  & $   42.70 $  &  $   4.63 \pm  0.22  $  &  $  -7.35 \pm  0.07  $  \\
  9    &  $  1.40 $  & $   41.58 $  &  $   0.57 \pm  0.10  $  &  $ -19.27 \pm  0.30  $  \\
 10    &  $  1.22 $  & $   38.52 $  &  $ -11.65 \pm  0.58  $  &  $  -6.02 \pm  2.20  $  \\
 11    &  $ -1.06 $  & $   31.90 $  &  $  -2.51 \pm  0.33  $  &  $ -33.15 \pm  0.21  $  \\
 12    &  $  2.98 $  & $   30.90 $  &  $   3.46 \pm  0.13  $  &  $ -42.85 \pm  0.14  $  \\
 13    &  $  1.53 $  & $   21.94 $  &  $  27.79 \pm  0.34  $  &  $-120.19 \pm  0.17  $  \\
 14    &  $  0.17 $  & $   17.21 $  &  $-189.07 \pm  0.14  $  &  $ -98.55 \pm  0.13  $  \\
 15    &  $  3.68 $  & $   14.88 $  &  $   4.39 \pm  0.19  $  &  $ -41.96 \pm  0.17  $  \\
 16    &  $  1.71 $  & $   13.24 $  &  $  30.20 \pm  0.44  $  &  $-119.75 \pm  0.14  $  \\
 17    &  $ -1.11 $  & $   11.42 $  &  $ -96.23 \pm  0.93  $  &  $ -16.53 \pm  0.46  $  \\
 18    &  $ -0.89 $  & $   11.22 $  &  $ -92.22 \pm  0.25  $  &  $ -15.88 \pm  0.27  $  \\
 19    &  $  4.60 $  & $    9.21 $  &  $   3.61 \pm  0.13  $  &  $ -32.36 \pm  0.26  $  \\
 20    &  $ -2.12 $  & $    8.94 $  &  $-100.14 \pm  0.14  $  &  $ -36.31 \pm  0.22  $  \\
 21    &  $  0.78 $  & $    8.25 $  &  $ -18.97 \pm  0.22  $  &  $  -6.01 \pm  0.18  $  \\
 22    &  $  2.80 $  & $    7.67 $  &  $   5.57 \pm  0.05  $  &  $ -21.12 \pm  0.06  $  \\
 23    &  $ -1.63 $  & $    6.77 $  &  $-110.54 \pm  0.40  $  &  $ -45.94 \pm  0.32  $  \\
 24    &  $  3.16 $  & $    6.77 $  &  $   3.03 \pm  0.18  $  &  $ -35.91 \pm  0.19  $  \\
 25    &  $  1.49 $  & $    6.69 $  &  $  32.92 \pm  0.16  $  &  $-121.14 \pm  0.15  $  \\
 26    &  $ -0.62 $  & $    6.21 $  &  $-125.04 \pm  0.27  $  &  $ -61.38 \pm  0.45  $  \\
 27    &  $  0.70 $  & $    5.45 $  &  $   1.96 \pm  0.09  $  &  $ -18.21 \pm  0.12  $  \\
 28    &  $  1.71 $  & $    5.43 $  &  $   7.74 \pm  0.20  $  &  $ -15.22 \pm  0.18  $  \\
 29    &  $ -1.46 $  & $    5.22 $  &  $  -0.06 \pm  0.19  $  &  $ -38.88 \pm  0.36  $  \\
 30    &  $ -1.55 $  & $    4.99 $  &  $   3.52 \pm  0.21  $  &  $ -40.09 \pm  0.11  $  \\
 31    &  $  1.84 $  & $    4.28 $  &  $  35.66 \pm  0.15  $  &  $-145.66 \pm  0.15  $  \\
 32    &  $  2.15 $  & $    3.93 $  &  $   2.46 \pm  0.30  $  &  $ -46.93 \pm  0.23  $  \\
 33    &  $  1.53 $  & $    3.79 $  &  $   7.09 \pm  0.17  $  &  $ -97.93 \pm  0.19  $  \\
 34    &  $  1.75 $  & $    3.72 $  &  $   4.15 \pm  0.23  $  &  $ -24.19 \pm  0.37  $  \\
 35    &  $  1.66 $  & $    2.85 $  &  $   5.91 \pm  0.21  $  &  $ -24.12 \pm  0.12  $  \\
 36    &  $  4.08 $  & $    2.83 $  &  $   2.93 \pm  0.12  $  &  $ -32.09 \pm  0.16  $  \\
 37    &  $  1.62 $  & $    2.78 $  &  $   7.74 \pm  0.29  $  &  $ -24.47 \pm  0.14  $  \\
 38    &  $  3.02 $  & $    2.21 $  &  $  16.56 \pm  0.15  $  &  $ -22.54 \pm  0.18  $  \\
 39    &  $  1.53 $  & $    2.07 $  &  $   9.77 \pm  0.53  $  &  $ -24.18 \pm  0.26  $  \\
 40    &  $ -1.37 $  & $    1.73 $  &  $-127.95 \pm  0.17  $  &  $ -58.28 \pm  0.17  $  \\
 41    &  $  2.32 $  & $    1.41 $  &  $  11.30 \pm  0.24  $  &  $ -21.75 \pm  0.14  $  \\
 42    &  $ -0.93 $  & $    1.33 $  &  $-129.40 \pm  0.33  $  &  $ -60.45 \pm  0.30  $  \\
 43    &  $  2.19 $  & $    1.33 $  &  $   5.98 \pm  0.29  $  &  $ -29.82 \pm  0.35  $  \\
 44    &  $  1.88 $  & $    1.09 $  &  $  27.57 \pm  0.23  $  &  $-147.83 \pm  0.21  $  \\
 45    &  $  2.23 $  & $    0.96 $  &  $   9.26 \pm  0.31  $  &  $ -53.23 \pm  0.39  $  \\
 46    &  $ -1.50 $  & $    0.95 $  &  $ 171.49 \pm  0.63  $  &  $  45.45 \pm  0.76  $  \\
 47    &  $  0.04 $  & $    0.85 $  &  $ -43.98 \pm  0.41  $  &  $   8.37 \pm  0.18  $  \\
 48    &  $ -1.76 $  & $    0.62 $  &  $   5.16 \pm  0.21  $  &  $ -51.65 \pm  0.23  $  \\
 49    &  $ -3.30 $  & $    0.56 $  &  $   8.15 \pm  0.25  $  &  $ -72.68 \pm  0.30  $  \\
 50    &  $  2.50 $  & $    0.47 $  &  $  35.80 \pm  0.29  $  &  $ -75.53 \pm  0.24  $  \\
 51    &  $ -2.82 $  & $    0.33 $  &  $  10.44 \pm  0.41  $  &  $-183.21 \pm  0.46  $  \\
 52    &  $ -3.04 $  & $    0.32 $  &  $  10.39 \pm  0.29  $  &  $-181.65 \pm  0.38  $  \\

\enddata

\tablecomments{Maser features detected within $1''$ from G009.62+00.20\,E listed by
decreasing brightness (Column~1). OH maser features are listed for each circular polarization
component. Columns~2 and~3 report the LSR velocity and brightness of the brightest spot of each
feature. Columns~5 and~6 give the relative centroid position of each feature in the east and north directions,
respectively. Reference absolute positions are reported in Table~\ref{obstab}. The uncertainties give the
intensity-weighted standard deviation of the spots' distribution within a given feature.}

\end{deluxetable}

\end{document}